\documentclass[manuscript,screen]{acmart}

\AtBeginDocument{%
  \providecommand\BibTeX{{%
    \normalfont B\kern-0.5em{\scshape i\kern-0.25em b}\kern-0.8em\TeX}}}

\copyrightyear{2020}
\acmYear{2020}
\setcopyright{acmlicensed}\acmConference[PEARC '20]{Practice and Experience in Advanced Research Computing}{July 26--30, 2020}{Portland, OR, USA}
\acmBooktitle{Practice and Experience in Advanced Research Computing (PEARC '20), July 26--30, 2020, Portland, OR, USA}
\acmPrice{15.00}
\acmDOI{10.1145/3311790.3396641}
\acmISBN{978-1-4503-6689-2/20/07}
\usepackage{listings}

\begin{document}

\title{Using Containers to Create More Interactive Online Training and Education Materials}

\author{Brandon Barker}
\email{brandon.barker@cornell.edu}
\orcid{0000-0001-5732-9550}
\author{Susan Mehringer}
\email{shm7@cornell.edu}
\orcid{0000-0002-2437-5494}
\affiliation{%
  \institution{Cornell University Center for Advanced Computing}
  \streetaddress{Frank H. T. Rhodes Hall}
  \city{Ithaca}
  \state{New York}
  \postcode{14853}
}

\renewcommand{\shortauthors}{Barker and Mehringer}

\begin{abstract}
Containers are excellent hands-on learning environments for computing topics because they are customizable, portable, and reproducible.  The Cornell University Center for Advanced Computing has developed the Cornell Virtual Workshop in high performance computing topics for many years, and we have always sought to make the materials as rich and interactive as possible.  Toward the goal of building a more hands-on experimental learning experience directly into web-based online training environments, we developed the Cornell Container Runner Service, which allows  online content developers to build container-based interactive edit and run commands directly into their web pages. Using containers along with CCRS has the potential to increase learner engagement and outcomes.
\end{abstract}




\maketitle

\section{Introduction}

\subsection{Background}
  Online training materials must be more than glorified documents to engage and teach learners.  Learning by doing and experimenting is critical to gain understanding and skills in computing and programming.  Integrating computing environments directly into online training web pages removes real and perceived access barriers.  While this capability exists for specific environments, e.g. Jupyter Notebook, we sought to develop a more flexible tool to support learning to code and the understanding of concepts that cover the full array of large-scale computing topics.  The tool we originally envisioned and developed, the Cornell Job Runner Service (CJRS), has been redesigned to use containers and now exists as the Cornell Container Runner Service (CCRS).  Our intention is to implement CCRS widely in our online materials (Cornell Virtual Workshop), and then  offer CCRS to the broader HPC training and education community as an API.
  
\subsection{Containers}

Containers \cite{CornellVirtualWorkshopContainerswithCR-2020-05-13},
as the name implies, allow packaging an application or environment in
a portable way - a desirable property to have for integrating computational material
into an educational service shared across multiple courses and instructors.
In particular, containers are a form of OS-level virtualization
where a single OS kernel can
host multiple isolated environments. They are lighter weight than using
hypervisor-based virtual machines  wherein each VM runs a separate kernel,
but comes with the obvious restriction that all containers on the host
must use the same OS kernel and the same version of the
kernel. In many cases, this is an acceptable
trade off, and while the roots of container technology originated in
the likes of Solaris
Zones \cite{10.5555/1052676.1052707} and
FreeBSD Jails \cite{Kamp00jails:confining},
the prevalence of Linux on commodity cloud hardware has
helped to create a convergence of Linux
containerization technology.

While the above definition remains true, more recently
Docker \cite{Chamberlain2014} has  popularized the use of Linux containers in 
many communities and
 expanded, at least colloquially, the definition
of a container. The definition now includes the ability
to distribute and deploy applications
with minimal configuration: i.e., everything is self-contained within
the
container itself. Singularity \cite{10.1371/journal.pone.0177459},
another Linux container technology, targeted HPC users by inverting
security concerns and requiring the user to grant an application
container access to all of the user's files rather than running a
service as the \verb!root! user, as is the case with Docker. In so doing,
many of the traditional notions of a container were further broken
down, though Singularity has optional parameters
to enable isolation
\cite{RunningServicesSingularitycontainer30documentation-2020-01-13}.

Other container technologies exist for Linux as well, including that which is
provided by a core service of most Linux distributions,
systemd \cite{systemdnspawn-2020-02-05}.
However, \verb!systemd-nspawn! containers typically do not include the more
modern connotation of a container being a packaged application. But,
it has been used by other technologies for this purpose, such as
nix-containers \cite{NixOSmanual-2020-02-06_containers}, a
container technology for the
NixOS \cite{dolstra_loh_pierron_2010,10.1145/3152493.3152556} Linux
distribution that allows packages to be shared from the host's
package-store.

\subsection{Cornell Job Runner Service (CJRS)}

CJRS (Cornell Job Runner Service \cite{10.1145/2792745.2792765})
is the predecessor to CCRS (Cornell Container Runner Service) but has been largely redesigned
and rewritten. CJRS employed the Slurm Workload Manager \cite{10.1007/10968987_3}.
and relied on extensions to Slurm which were non-trivial to install
and sometimes error-prone. CJRS was developed
just before containers \textit{as an environment packaging technology} became prevalent,
and once containers in this form became widespread, it was clear that pairing
CJRS with container technologies would be extremely useful by providing
the same
front-end features and course-integration flexibility as CJRS with 
the  flexibility and low maintenance cost of containers.

\section{Cornell Container Runner Service (CCRS)}
\subsection{Overview}
CCRS provides several interactive job-runner elements that instructors
can use out of the box, e.g. an editors coupled to code compilers or code runners
and one-shot, in-line commands. Coupled with its support of a popular HPC
container type and a container type with low-disk utilization, CCRS provides
a low-cost, cloud-friendly solution for HPC and CS instructors who want to integrate
their exercises into an online platform or site of their choice.
CCRS takes the ideas and some of the architecture of CJRS and replaces the job
lifecycle management of Slurm with  containers. While this involves
managing life-cycle explicitly, this was on the same order of effort needed to tailor
Slurm to manage jobs on virtual nodes in CJRS.
CCRS provides an API that instructors can use to incorporate their examples,
as well as allowing the creation of
exercises in novel formats that  go beyond what we envisaged.

\subsection{CCRS Features}

\subsubsection{CCRS Job Types}

CCRS has several different \textit{job types} that may be employed for exercises
and examples. Currently, the two prominent job types are 
\textit{one-shot commands} and an editor-based \textit{job launch application}.
The one-shot command allows a single command to be entered or modified
and run under whatever language the instructor has specified. For job-launch
applications, typically an editor is employed. The instructor can add various
actions to the editor as desired, such as "run" or "compile and run."
Interactive shell support is also planned. In the case of both the job launch
application and interactive shell, CCRS has taken the approach of using
third-party libraries, such as Ace editor \cite{AceTheHighPerformanceCodeEditorfortheWeb-2020-02-01},
which includes syntax highlighting
for many languages; CJRS took the alternative
approach of using minimal code that had few features but was
also smaller in size. If the demand arises, we will integrate these
lighter-weight alternatives into CCRS as well.

While these features encompass what was originally provided by CJRS, additional features are under consideration, including graphical and file-based interaction, where images
generated by a job may be previewed and images and other generated files may be
downloaded by the user.

\subsubsection{Container Types} \label{container_types}

Currently, CCRS supports Singularity and \verb!systemd-nspawn! containers;
we anticipate supporting
others in the future, including Docker, though Docker has lower
priority due to current and historical security
issues \cite{DBLP:journals/corr/abs-1804-05039}.
Nix containers are also employed
via \verb!systemd-nspawn! support; they provide a light-weight
container approach when building up specific environments for instructors
since there is no package storage overhead. Singularity is
also supported primarily due to its popularity in HPC and the number of
instructors who may have an existing Singularity container they wish to use.
Generally, the overhead of adding a new container type should not be
that large, as CCRS abstracts the container management lifecycle as much
as possible.

\subsubsection{Container Lifecycle}

\begin{figure*}
  \centering
  \includegraphics[trim=50 200 250 50,clip,width=\textwidth]{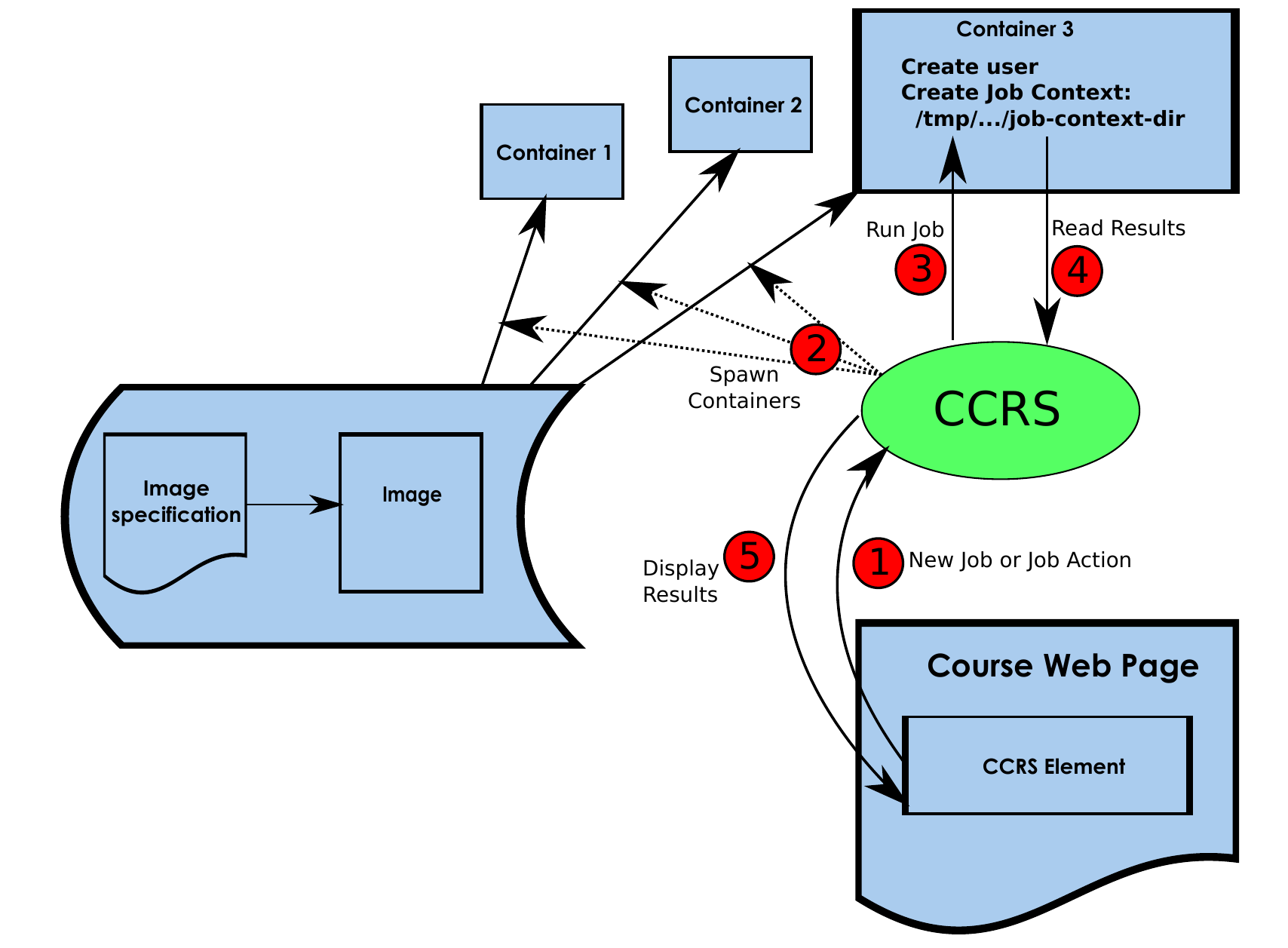}
  \caption{\textbf{CCRS container lifecycle}. Not shown is the container-type-dependent handling of garbage collection for of job contexts, users,
  and containers.}
\label{fig:containerLifecycle}
\end{figure*}

A diagram of the container and job management approach taken by CCRS is shown in
Figure \ref{fig:containerLifecycle}.
Generally, containers are environments determined by an \textit{image},
which is a snapshot of a filesystem built from a specification
(e.g. Dockerfile, Singularity definition file, or Nix expression). Multiple containers
can then be launched from the same image, which in CCRS can be done on
a per-user basis or a per-job basis.

For Singularity, CCRS will create a host user
when valid login credentials are provided from the training site.
The user's login will be specific to
the course site the user is coming from, rather than global across
all CCRS courses, though this should not generally be relevant to the student.
To achieve true containerization, we pass the
\verb!--containall! option to \verb!singularity exec!, which will then
contain not only file systems, but also PID, IPC, and environment
\cite{singularityexecSingularitycontainer35documentation-2020-01-13}.
If Docker support were to be added, it would likely follow a model similar 
to this, but with the added need to garbage-collect image layers generated by
running containers. In the case of Nix containers, images are not employed,
but we still make use of the
abstractions; instead, all jobs for a specific container specification are run in
the same container by default, but as different users created within the container
(again, based on the supplied login credentials).
The container itself acts as a sort of image, and jobs get user accounts in that
container rather than specific container invocations, which is much 
lighter weight.
In this regard, we are again using the more modern connotation of containers
as application or environment packaging platforms rather than as a means of
isolated process-sets. Using the \verb!machinectl clone!
could achieve process isolation, however, we believe 
this is largely overkill for our use case.

Within CCRS, each job is given a unique ID (the \textit{job ID}). 
The job ID is in one-to-one correspondence with the \textit{job context} 
(which is taken from CJRS); the job context is
essentially the path where the job's user can write files
and includes the identifier as part of the path. The job contexts are not
garbage-collected when a job completes, though eventually they can be (for
instance, by default we have configured CCRS to use a folder under
\verb!/tmp! \cite{318tmpTemporaryfiles-2015-05-20}
which may have site-specific persistence policy, e.g., files are cleared
after a reboot or after a certain period). Instead, they are kept
around to diagnose any possible issues that may arise.
Sometimes the ID --- and thus job context --- will persist across multiple user
actions, as is the case for editor-based examples. On the
other hand, for single commands, a new job id and context will be created 
each time, though the CCRS API (see below) allows the instructor some flexibility
in this regard.

Since the job context resides on the host filesystem to be managed by CCRS
more readily, all container types supported by CCRS must support a way to mount
directories on the host. Since this is a fairly standard feature, all container
types we've come across so far support some version of host-filesystem mounts.

\subsubsection{Implementation Notes}

The CCRS web service is implemented in Scala and Scala.js
(which compiles Scala to JavaScript), allowing a uniform language
and RPC (remote procedure call) API for the front-end and back-end; there is no need
to use JavaScript directly, though JavaScript libraries are used in some cases, like
Ace editor. As much as possible,
the design has been purely functional and uses the ZIO \cite{zioGitHub-2020-02-06}
library to track effects and errors, and to simplify concurrency,
which improves code reliability and allows for simpler refactors
\cite{HappyeyeballsalgorithmusingZIOSoftwareMillTechBlog-2020-02-06,WhyweloveScalaatCourseraCourseraEngineeringMedium-2020-02-06}.

Logging has been implemented for
CCRS so that tracking occurs for malicious users' jobs or
any errors that occur in the container lifecycle, either in the user's code
or in CCRS itself. In the case of errors or normal actions occurring in
the CCRS life cycle, the log contains the job ID, so that the job's
context may be later associated with the relevant logs.

Instructor's sites are considered privileged, in that they may create users with a
site-specific user prefix and launch jobs. If any aberrant behavior is noticed from
an instructor's site, the site can be temporarily disabled until the issue is resolved.

\subsubsection{The CCRS API} \label{ccrs_api}

Instructors may use one of two CCRS APIs to
add this functionality to their sites: the JavaScript API or the Scala.js API. We
anticipate almost all users will use the JavaScript API, as most users are
more familiar with JavaScript, and it requires no additional infrastructure.
The JavaScript API simply wraps the Scala.js API, which is the native implementation
of the API. The Scala.js API is still helpful for debugging and development of the
API, as examples can be directly included using dynamically-generated HTML in
the CCRS product. If an example is written in Scala.js, a
corresponding JavaScript example can then be added.

To use the API, the page must load the requisite JavaScript code, optionally
including \verb!ace.js! if an editor-based example is used on the page:

\begin{lstlisting}[language=HTML, basicstyle=\footnotesize]
<script type="application/javascript" src="http://w.x.y.z:port/ace/ace.js" charset="utf-8"></script>
<script type="application/javascript" src="http://w.x.y.z:port/target/web-client-jsdeps.js"></script>
<script type="application/javascript" src="http://w.x.y.z:port/target/web-client-opt.js"></script>
\end{lstlisting}

Implementation of a one-shot command in a web page might look like this:

\begin{lstlisting}[language=HTML, basicstyle=\small]
<h2>Free-form single-command input</h2>
<input type="text"
       placeholder="Enter a command:"
       value="pwd"
       onkeydown="oneShotHandler(event)" />
</body>
<div id="one-shot-demo"></div>
\end{lstlisting}

First, the \verb!input! form is created. Then, a \verb!div! element is created
where CCRS will display output information to the user, such as standard output
and standard error.

Next, the metadata and input form handler are specified in a \verb!script!
element:

\begin{lstlisting}[language=HTML, basicstyle=\small]
<script type="application/javascript">
  var ccrsApiNamespace =
    "org.xsede.jobrunner.model.ModelApi";
  var pythonExampleMetaJson = {
    "$type": ccrsApiNamespace + ".SysJobMetaData",
    "shell": ["bash"],
    "containerType": {
      "$type":  ccrsApiNamespace + ".Singularity"
    },
    "containerId": [],
    "image": ["vsoch-master-latest.simg"],
    "binds": [],
    "overlay": [],
    "user": "ccrsdemo",
    "address": [],
    "hostname": [],
    "url": window.location.href
  };
  var pythonExampleMeta = 
    CCRS.sysJobMetaData(pythonExampleMetaJson);
  var oneShotId = CCRS.makeJobId();
  var oneShotCommand = CCRS.makeOneShotCommand(
    document.getElementById("one-shot-demo")
  );
  var oneShotHandler = CCRS.makeCmdHandler(
    oneShotCommand,
    pythonExampleMeta,
    oneShotId
  );
</script>
\end{lstlisting}

The metadata specification includes important coordinates such as the
system shell to use in the container, container type, container id
(potentially used when resuming an existing job, such as in a guided session),
container image, and  user. In practice, some fields will be filled
in by the CCRS client code automatically, such as the \verb!address!,
\verb!hostname! and  \verb!url!.

After the metadata specification, a command output view is created with
\verb!oneShotCommand!, which takes the \verb!div! element mentioned above as an argument
so that it may write job-related information to that \verb!div! element.
This command handle is then passed to the handler creation \verb!makeCmdHandler!
utility function. 
Note that creating a job launch application using an
editor pane is very similar, except that in addition, another \verb!script!
element can be used to embed code in any given language. Often, the same metadata
can be shared between examples, so it need only be defined once if the same
container is used throughout a site. Detailed
examples are available in the CCRS repository.

\begin{figure*} 
\includegraphics[width=8cm]{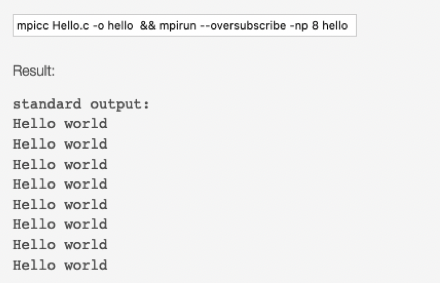}
\caption{Implementation of a one-shot command application, showing MPI compile and run of Hello.c}
\label{fig:oneShot}
\end{figure*}

An example of a one-shot command application, i.e. an instructional web page with an input box to run a single command, is shown in Figure \ref{fig:oneShot}.  This application is useful when the instructor wants to walk the student through a series of commands, or to allow the student to experiment with a list of commands.  This could be used when learning a new operating system, compiling code, or for tasks such as file transfer.  The example shown demonstrates compiling, then running, a simple MPI program, \verb!Hello.c!; in this case we are using \verb!&&! to enter two commands into the one-shot command box.   

\begin{figure*} 
\includegraphics[width=8cm]{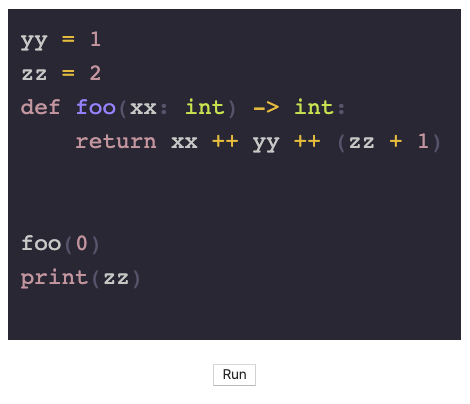}
\caption{Implementation of a job launch application, showing an edit box and run command embedded in the web page}
\label{fig:jobLaunch}
\end{figure*}

 Figure \ref{fig:jobLaunch} is an example of a job launch application, i.e. an instructional web page with an embedded editor page and code run capability.  This example is from instruction on functional programming.  The instructor can supply  code which may be complete, incomplete, or intentionally broken.  The student can experiment with the code by editing and running in line, learning by experimentation and observation, and following the instructor's written guidance. 

In practice, an instructor who wants to implement custom behavior
could look at the source of  \verb!makeOneShotHandler! and add custom
modifications to create their own job type to interact with CCRS. An example
of this is a custom guided-session. This admittedly requires
some knowledge of JavaScript, but as the usage of CCRS increases,
we plan to create a higher-level API,
primarily for JavaScript users, that provides commonly used job types out of
the box, and with less boiler-plate than what is seen above.
This would essentially remove the need to
specify the metadata structure except for a few necessary fields.

\subsubsection{Deployment and Maintenance}

We believe CCRS is best used as a cloud virtual machine. In the case of a compromised
VM, we can quickly spin up another VM and
grant it the floating IP used by the compromised system. At the same time,
we can perform a post-mortem check of the compromised instance while monitoring
the live instance for aberrant behavior. Currently, we host an instance on Cornell's
Red Cloud \cite{10.1145/3355738.3355755}.
A complete template of the system is made available as a
\verb!configuration.nix! file, allowing the system to easily be rebuilt
without distributing and hosting a costly image, or allowing further customization
of the system for parties interested in hosting their own CCRS installation.

\subsubsection{Architecture-specific code}

In rare cases, instructors may wish to go fairly low-level and discuss
topics that are specific to certain architectures. While there is a limit
to what can be achieved, if emulators for the platform are available
(e.g. POWER9 \cite{OpenPOWERFoundationIntroducingIBMPOWER9FunctionalSimulator-2020-02-06}
or Intel AVX-512 \cite{IntelSoftwareDevelopmentEmulatorIntelSoftware-2020-02-06}),
then a container that houses the emulator
and abstracts away usage of the emulator could be constructed. Since
most learning exercises are not intended to be compute-intensive, using an
emulator should not typically be a problem. Furthermore, this makes the course
materials still accessible and interactive after a particular resource
is decommissioned or, in rare cases, an emulator may be provided
before the HPC resource is brought online, allowing instructors and students
a head start on working with the architecture.

\section{Comparison with Related Work}

\subsection{Notebooks}

\subsubsection{Jupyter and iPython Notebook}

A tool that many in education are familiar with is Jupyter, or its predecessor,
iPython Notebook \cite{10.1145/3219104.3219162,10.1145/3219104.3219122}.
Jupyter can certainly be an effective tool for learning
in the classroom \cite{10.1145/3093338.3093370}
or online. Both CCRS and Jupyter aim to
provide a simple and interactive experience for the student. However, there are
significant differences in Jupyter and in CCRS and, in many cases, CCRS may be
the ideal solution, especially when developing custom courses or
training materials. CCRS caters to specialization in two regards: user interface
and customized containers.

Jupyter presents a convenient notebook-style interface, where results for an
in-depth demo can be evaluated, modified, and evaluated again in an iterative
fashion. We feel that this is great for a tutorial and can also be useful for
some parts of a course, particularly if the course is running one of the
standard kernels. While Jupyter's notebook-style interface is a combination of
an editor and REPL, CCRS supports several more standard UI elements that can
also be customized by an instructor if desired, as described in 
Secs~\ref{container_types}~\&~\ref{ccrs_api}. 
It is completely up to the instructor which of the UI
elements to use for each example. JupyterLab \cite{jupyterlabGitHub-2020-02-06}
presents a customizable MATLAB-like user interface, which is likely suitable for
data exploration and analysis but still doesn't offer embeddable and
customizable UI elements.

Unlike Jupyter, CCRS will require the instructor to use a very simple JavaScript
API to create interactive examples in their web pages. We don't expect this to
be much of a burden, especially for what we anticipate to be our most typical
examples: an editor coupled with some action (e.g. run for Python, or compile
and run for C++). We plan to include ample examples so instructors can get up
and running with minimum effort.

As for customized environments, Jupyter has made great strides over the original
IPython, but this flexibility involves installing a custom kernel (which is not
a standardized procedure), and in some cases, becoming familiar with
creating a custom kernel, which is much more involved than just creating a
container. An example might be creating a kernel that supports more than one
language, which while technically possible, seems to be an ongoing issue as of
this writing \cite{JupyterLabIssue2815-2020-02-06}, though
BeakerX \cite{BeakerX-2018-10-17} appears to address the issue for some
languages. CCRS supports a variety of container formats.

At the time of this writing, there is no Jupyter kernel for running MPI with C,
which is one of the prerequisites for the MPI course and online book which will
utilize CCRS. A goal of CCRS is to provide tests and autograding for such
environments. Autograding can be accomplished via nbgrader
\cite{nbgradernbgrader061documentation-2019-11-07}
in Jupyter, but it involves the
installation of third-party software with Jupyter, which in our experience, can
result in significant maintenance overhead and does not always proceed smoothly.
In contrast, CCRS is intended to be a single self-contained system; it will be
open-source, so extensions can easily be added directly into the codebase, either
by forking or hopefully by passing the core developers' review and becoming
integrated into the standard CCRS distribution.

Neither Jupyter nor CCRS inherently addresses the issue of how to install the
software being used in a course, which is often a valuable skill to have.
However, if we take, for instance, MPI, many users will never need to install or
configure MPI themselves as they may just be running code on HPC clusters where
they have access. Other users will want to learn more about installing their
software stack so they can have it on their personal
system. For this reason, we  advocate publishing all container types so that
users can easily use the container directly, or inspect the container
specification file so they can see how the environment was constructed; all
courses in the pilot will have their container specifications made public both
through the CCRS GitHub repository and through the associated course.

\subsubsection{Apache Zeppelin}

Apache Zeppelin \cite{10.1145/3219104.3229288} is similar in its user interface
to Jupyter, and is focused on data science and primarily is known for Python and
Scala notebooks, but much like with Jupyter's kernels, a custom Interpreter
needs to be created when adding a new programming language or environment (like
Apache Spark).

\subsection{Other Training Formats}

\subsubsection{MOOCs}

MOOCs (Massive Open Online Courses) often have their own systems in
place for running exercises. While we are
not competing with MOOCs directly, CCRS provides a technology that could be
used by MOOCs under the hood, just as it could be integrated with almost any
site. While many computer science MOOCs exist, some HPC examples include
\textit{Fundamentals of Parallelism on Intel Architecture}
\cite{52MessagePassingInterfaceClustersandMPICoursera-2020-02-06}
and \textit{Future Learn: Supercomputing} \cite{SupercomputingOnlineCourse-2020-02-06}.

\section{Future Work}

Additional features are planned for CCRS, including a shell (terminal) emulator,
a high-level API to reduce boilerplate for common examples, improved security
features, and the ability to interact with images and other files that may
be part of an example. Security features would include items such as
time-to-live, CPU, filesystem, and memory restriction  policies, as well as 
run-time container checks and improved logging. Once the APIs are solidified 
and known security features completed, we plan to release CCRS as open-source,
allowing CCRS to be hosted at other sites as well as on systems with exotic hardware.

\section{Acknowledgments}

This work was in part supported by the Extreme Science and Engineering Discovery Environment (XSEDE), which is supported by National Science Foundation grant
number ACI-1548562.

We also thank Victor Eijkhout at the Texas Advanced Computing Center for being an early user of CCRS in his course.

\bibliographystyle{ACM-Reference-Format}
\bibliography{ccrs-base}











\end{document}